\documentclass[apj]{emulateapj}
\usepackage{graphicx}
\usepackage{natbib}
\usepackage{amssymb}

\begin{document}

\title{Unveiling the universality of I-Love-Q relations }
\author{Y.-H. Sham\footnote{Email: yhsham@phy.cuhk.edu.hk},
T.~K. Chan\footnote{Present address: Department of Physics,
University of California at San Diego, 9500 Gilman Drive, La
Jolla, CA 92093, USA; Email: chantsangkeung@gmail.com},   
L.-M. Lin\footnote{Email: lmlin@phy.cuhk.edu.hk}, and 
P.~T. Leung\footnote{Email: ptleung@phy.cuhk.edu.hk} }
\affiliation{Department of Physics and
Institute of Theoretical Physics, The Chinese University of Hong
Kong, Hong Kong SAR, China}

\date{\today}
\begin{abstract}
The recent discovery of the universal I-Love-Q relations connecting the 
moment of inertia, tidal deformability, and the spin-induced quadrupole 
moment of compact stars is intriguing and totally unexpected. In this paper,
we provide numerical evidence showing that the universality can be attributed 
to the incompressible limit of the I-Love-Q relations. The fact that modern 
equations of state are stiff, with an effective adiabatic index larger 
than about two, above the nuclear density range is the key to establishing
the universality for neutron stars and quark stars with 
typical compactness from about 0.1 to 0.3. 
On the other hand, the I-Love-Q relations of low-mass neutron stars near 
the minimum mass limit depend more sensitively on the underlying equation of 
state because these stars are composed mainly of softer matter at low 
densities. However, the I-Love-Q relations for low-mass quark stars can still 
be represented accurately by the incompressible limit. 
We also study the I-Love relation connecting the moment of inertia and tidal 
deformability analytically in Newtonian gravity and show why the I-Love-Q 
relation is weakly dependent on the underlying equation of state and can be 
attributed to its incompressible limit. 
\end{abstract}

\maketitle

\section{Introduction}
\label{sec:intro}

Neutron stars (NSs) are prime examples of physical systems that require an
understanding of both strongly interacting many-body nuclear physics and 
general relativity (GR). 
With the density in their cores reaching a few times the normal nuclear 
density, that is, the regime where theoretical calculations for dense nuclear 
matter
are poorly constrained, the equation of state (EOS) in the cores of NSs 
is not well understood. Different EOS models would generally predict  
quite different physical quantities related to NSs. 
From a nuclear physicist's point of view, it would thus be interesting 
to use the observed physical quantities of NSs to place constraints
on EOS models. 
For example, the discovery of NSs with masses of about $\sim 2M_\odot$ 
\citep{Demorest:2010p1081,Antoniadis:2013p6131} would place a strong 
constraint on the underlying EOS model (see, e.g., \citet{Lattimer:2012p485} 
for discussions). 
A relativist, however, might be more interested 
in looking for EOS-insensitive relationships connecting the different physical 
quantities, since one could then use these relations to test the underlying
gravitational theory in the strong-field limit despite our ignorance of the 
supranuclear density EOS. 

It is well established that different EOS models would generally lead to
quite different stellar structure and global quantities of NSs, 
such as their masses and radii. It is thus quite surprising that 
approximately EOS-insensitive relations concerning various different physical 
quantities, such as the moments of inertia, compactness, and oscillation 
modes of nonrotating NSs, can indeed be found 
\citep{Bejger:2002p8392,Lattimer:2005p7082,Andersson:1998p1059,
Andersson:1996p20, Benhar1999:p797,Benhar:2004:p124015,Tsui:2005p151101,
Tsui:2005p1029,Lau:2010p1234}. 
More recently, the discovery of the so-called I-Love-Q relations in GR 
\citep{Yagi:2013long,Yagi:2013} has gained a lot of interest. 
The I-Love-Q relations connect the moment of inertia, the spin-induced 
quadrupole moment, and the tidal deformability of NSs with 
EOS-insensitive relations.

Since the original discovery by \citet{Yagi:2013long,Yagi:2013}, the 
investigation of the I-Love-Q relations has been extended to include rapid 
rotation. 
Initially, \citet{Doneva:1310.7436} used the rotation frequency to 
characterize rotation and found that the I-Q universal relation is broken.
However, more recent numerical works by 
\citet{Pappas:1311.5508} and \citet{Chakrabarti:1311.6509}, 
which are supported by 
the analytical study of \cite{Stein:2014p15}, found that the I-Q relation
remains approximately EOS-independent for rapidly rotating stars when 
rotation is characterized by a suitable dimensionless parameter. 
Besides the extension to rapid rotation, the effects of magnetic field 
\citep{Haskell:1309.3885} and tidal deformation in NS-NS inspirals 
\citep{Maselli:2013} on the universal relations have also been studied.
More recently, universal relations connecting higher multipole moments 
have also been investigated \citep{Pappas:1311.5508,Stein:2014p15,
Yagi:2014p124013,Chatziioannou:1406.7135}. 
As proposed in \citet{Yagi:2013long,Yagi:2013}, one interesting 
astrophysical application of the I-Love-Q relations is to use them to break 
the degeneracy between the NS quadrupole moment and the NS's individual 
spins in the gravitational-wave analysis for an inspiraling NS binary.

From a relativist's point of view, as discussed above, the I-Love-Q relations 
are interesting because they are insensitive to EOS models to within 1\%, 
and hence they could in principle be used to test GR by comparing the 
relations with the relevant observed quantities.\footnote{Note that 
approximately EOS-independent relations connecting the $f$-mode oscillation 
frequency to the mass and moment of inertia of NSs accurate to within 1\% 
level have also been found \citep{Lau:2010p1234}.}
Furthermore, it would be interesting to study whether similar universal 
relations exist in alternative theories of gravity. In the original work 
of \citet{Yagi:2013long,Yagi:2013}, besides discovering the I-Love-Q relations 
in GR, they also found similar universal relations in dynamical 
Chern-Simons gravity, although the relations differ from the GR ones. 
They further suggested that if the moment of inertia and the tidal 
deformability of the double-binary pulsar J0737-3039 can be measured 
to 10\% and 60\% accuracy, respectively, then the Chern-Simons theory can 
be constrained much better than current tests by six orders of magnitude.
Recently, the I-Love-Q relations were studied in Eddington-inspired
Born-Infeld gravity \citep{Sham:2014p66} and scalar-tensor theories 
\citep{Pani:1405.4547,Doneva:1408.1641}. 
It was found that the I-Love-Q relations in these theories agree with those 
in GR to within a few percent level.

The universal I-Love-Q relations are totally unexpected and the reason why 
these relations exist is not yet understood. In the original work,  
\cite{Yagi:2013long,Yagi:2013} suggested two possible reasons. 
The first reason is that the 
physical quantities considered depend mostly on the NS outer layer, where all
EOS models are similar. The second reason is that the relations approach the 
black hole limit as the NS compactness increases and hence the internal 
structure of NS becomes unimportant. 
More recently, \citet{Yagi:1406.7587} suggested that the 
isodensity contours of realistic NSs can be approximated by elliptical 
isodensity contours, and the assumption of the self-similarity of these 
surfaces plays a crucial role in the universality of I-Love-Q relations
(see Section~\ref{sec:isodensity} for a discussion).

In this paper, we propose that although realistic EOS models can differ 
from one another quite significantly, they are nevertheless already stiff 
enough in the nuclear density range to lead to the universality of the 
I-Love-Q relations. The stiffness of an EOS is measured by its 
effective adiabatic index $\Gamma$. 
While it is well known that $\Gamma$ generally depends quite sensitively 
on the density and varies from one EOS to another, the fact that modern 
realistic EOS models are typically stiff enough and have $\Gamma \gtrsim 2$ 
above the nuclear density range is enough to explain the universality of 
the I-Love-Q relations.
We will study analytically how the I-Love relation 
in Newtonian gravity depends on the stellar density profile, and hence the 
underlying EOS model, which is parameterized by a perturbation parameter 
$\delta$. We demonstrate that the prescribed density profile with the value
$\delta=1$ can yield the NS structure constructed by realistic EOS models 
very well. Furthermore, we show that the perturbation expansion point 
$\delta=0$, which corresponds to an incompressible stellar model (with 
$\Gamma = \infty$ formally), is a stationary point for the I-Love relation in 
the sense that the dependence of the relation on the stellar density profile 
is second order in $\delta$, and hence is weakly dependent on the EOS. 
Our analysis suggests that, in view of the I-Love relation, different 
realistic EOS models are generally stiff enough to be modeled well by a 
single incompressible EOS, and hence lead to an approximate universality.

The plan of this paper is as follows. In Section~\ref{sec:i-love-q}, we show 
numerically that the I-Love-Q relations obtained from realistic EOSs can be 
accurately represented by those of the incompressible EOS. 
We then provide a Newtonian analysis in Section~\ref{sec:Analy} to 
show that why this is the case for the I-Love relation. 
In Section~\ref{sec:breakdown}, we study when the universality of I-Love-Q 
relations would break down. 
Section~\ref{sec:isodensity} compares our finding with the recent work of 
\citet{Yagi:1406.7587} in which the elliptical isodensity approximation is 
suggested to play a crucial role in the universality. 
Finally, our conclusions are summarized in 
Section~\ref{sec:conclude}. Unless otherwise noted, we use geometric units 
where $G=c=1$.

\section{I-Love-Q relations}
\label{sec:i-love-q}

Here, we will study the I-Love-Q relations numerically 
in the slow-rotation approximation.
In GR, the moment of inertia $I$ of a star is obtained by 
expanding the expression $I = J/\Omega$ to first order in $\Omega$, where 
$J$ and $\Omega$ are the angular momentum and angular velocity of the 
star, respectively. The spin-induced quadrupole moment $Q$ characterizes 
the deformation of the star due to rotation and must be calculated at 
second order in $\Omega$.
The tidal deformability $\lambda$ measures the deformation of a 
star due to the tidal effect created by a companion star and is defined by 
$Q_{ij} \equiv - \lambda {\cal E}_{ij}$, where $Q_{ij}$ is the 
traceless quadrupole moment tensor of the star and ${\cal E}_{ij}$ is the 
tidal tensor that induces the deformation \citep{Flanagan:08p021502}. 
The methodologies to calculate $I$, $Q$, and $\lambda$ in GR 
can be found in \citep{Hartle:67p1005,Hartle:68:p907,Flanagan:08p021502,
Hinderer:08p1216,Damour:09p084035,Binnington:09p084018,Yagi:2013long,
Urbanec:13p1903}. 

\begin{figure}
  \centering
  \includegraphics*[width=7.5cm]{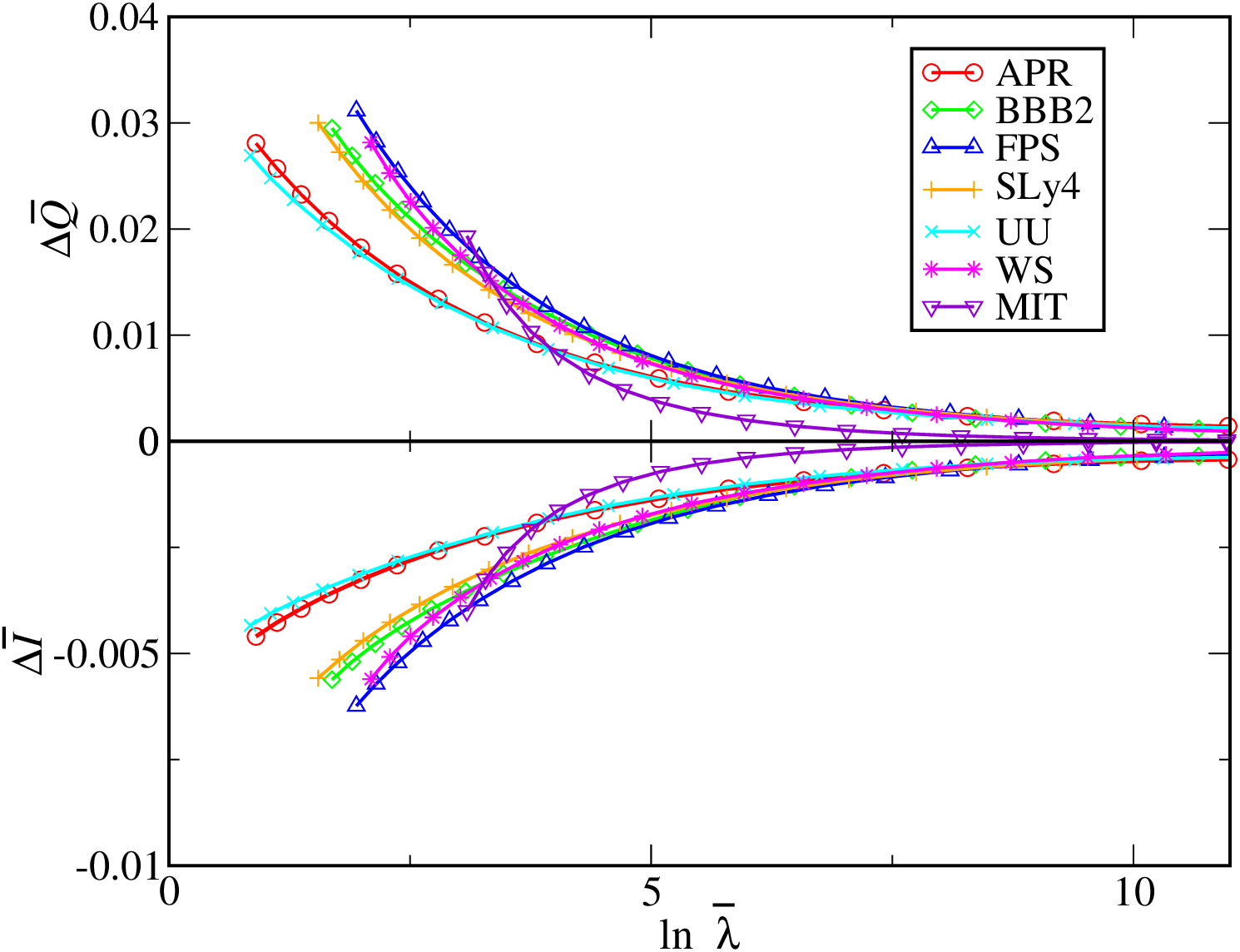}
  \caption{ $\Delta \bar Q$ (upper panel) and $\Delta \bar I$ (lower panel) 
are plotted against $\ln \bar \lambda$ for our chosen EOS models.  }
  \label{fig:i-love-q_real}
\end{figure}

The universal I-Love-Q relations \citep{Yagi:2013long,Yagi:2013} concern the 
dimensionless quantities  ${\bar I} \equiv I/M^3$, 
${\bar Q} \equiv - Q/(M^3 \chi^2)$, and
${\bar \lambda} \equiv \lambda / M^5$ (with $M$ and $\chi\equiv J/M^2$
being the mass and dimensionless spin parameter, respectively).  
The I-Love-Q relations are EOS-insensitive in the sense that if one plots 
${\bar I}$ against ${\bar \lambda}$ (and similarly ${\bar Q}$ against 
${\bar \lambda}$), then one would find that the results of the different
EOS models generally agree to a percent level (see, e.g., Figure 1 of 
\citet{Yagi:2013long}). 
In this work, we will consider the following nuclear matter EOS models: 
model APR \citep{Akmal:98p1804}, model BBB2 \citep{Baldo:1997p274}, 
model FPS \citep{Lorenz:93p379}, 
model SLy4 \citep{Douchin:2000p107}, model UU \citep{Wiringa:98p1010} and
model WS \citep{Lorenz:93p379,Wiringa:98p1010}. 
For comparison, we also consider quark stars (QSs) governed by the MIT bag 
model EOS with a bag constant of $B=70.2\ {\rm MeV\ fm}^{-3}$
\citep[see, e.g.,][]{Witten:1984p272,Alcock:1986p261}.

To motivate our analytic investigation of the I-Love relation in 
Section~\ref{sec:Analy}, 
which uses an incompressible stellar configuration as a background for 
perturbative expansion, here we first compare the I-Love-Q relations of 
realistic EOS models with those of an incompressible stellar 
model. In particular, we will use the results of an incompressible 
model as a benchmark to compare with those of realistic EOS models.
To achieve the comparison, for a fixed $\bar\lambda$ we define the 
fractional difference 
\begin{equation}
\Delta {\bar I} \equiv { {\bar I} - {\bar I}_{\rm incom}  \over 
{\bar I}_{\rm incom} } , 
\label{eq:delta_I} 
\end{equation}
where ${\bar I}$ and ${\bar I}_{\rm incom}$ are the scaled moments of 
inertia of a realistic EOS model and the incompressible model, 
respectively. Similarly, we define the fractional difference 
$\Delta {\bar Q}$ for the scaled spin 
induced quadrupole moment. In Figure~\ref{fig:i-love-q_real}, we plot 
$\Delta {\bar Q}$ (upper panel) and $\Delta {\bar I}$ (lower panel) 
against $\ln {\bar \lambda}$ using our chosen EOS models. 
Note that each curve terminates at the maximum mass limit for the 
corresponding EOS in the figure. 
It should be pointed out that $\bar \lambda$ decreases with increasing 
compactness. That is, general relativistic effects become 
important as $\bar\lambda$ decreases.
The compactness of NSs increases from about 0.07 to 0.3 as $\ln {\bar\lambda}$ 
decreases from 10 to about 1. Figure~\ref{fig:i-love-q_real} shows that the 
results of realistic EOS models generally agree very well 
with those of the incompressible model. 
In particular, even in the ultra-relativistic regime near the maximum mass 
limits (i.e., $\ln \bar \lambda$ in the range between one and two), the 
incompressible results still agree with realistic EOS results to about 3\% 
and 0.1\% for $\bar Q$ and $\bar I$, respectively.

Figure~\ref{fig:i-love-q_real} shows that the difference between the 
incompressible and realistic-EOS results generally varies with the EOS 
models according to their stiffness, as expected. 
For NS models, the APR and UU EOSs have maximum mass limits
at about $2.2 M_\odot$ and are stiff among the chosen EOS models. 
These two models have $\Delta \bar I$ at about 0.3\% for 
$\ln \bar\lambda = 2$. On the other hand, the FPS EOS is relatively soft 
(with a maximum mass $1.8M_\odot$) and the corresponding $\Delta\bar I$ 
increases to about 0.6\% for the same value $\ln \bar\lambda = 2$. 
We also see from Figure~\ref{fig:i-love-q_real} that the fractional 
differences for QSs governed by the MIT bag model are generally smaller 
than those of NSs for $\ln \bar\lambda$ larger than about five. 
This is expected as the density profile of QSs is rather uniform and can 
be modeled well by the incompressible stellar model. The deviations become 
comparable to those of NSs only near the maximum mass limit.

\begin{figure}
  \centering
  \includegraphics*[width=7.5cm]{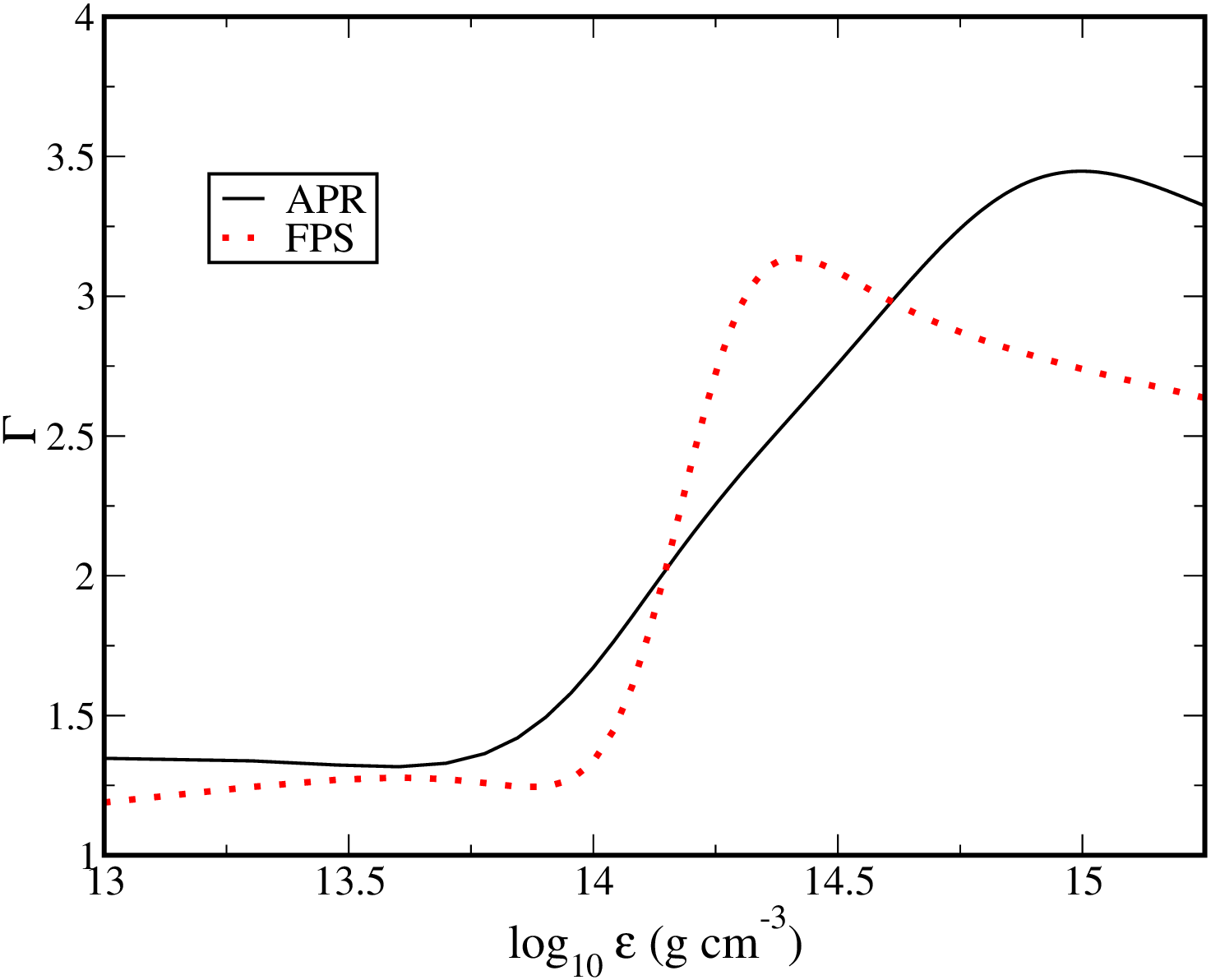}
  \caption{Effective adiabatic indices $\Gamma$ for the APR and FPS EOSs are
plotted against the energy density $\epsilon$.  }
  \label{fig:effective_gamma}
\end{figure}

\begin{figure}
  \centering
  \includegraphics*[width=7.5cm]{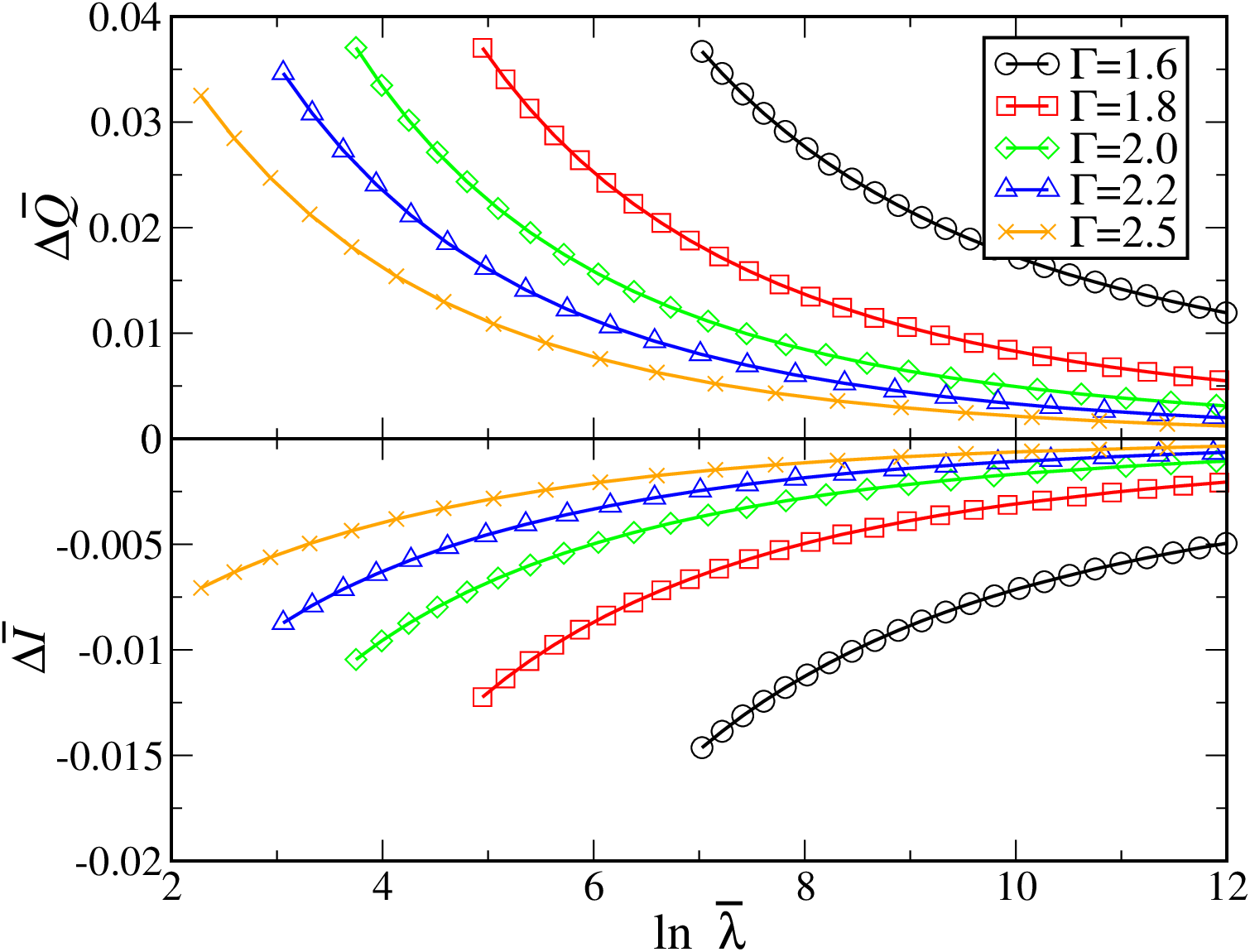}
  \caption{$\Delta \bar Q$ (upper panel) and $\Delta \bar I$ (lower panel) 
are plotted against $\ln \bar \lambda$ for polytropic models. }
  \label{fig:i-love-q_poly}
\end{figure}

The stiffness of an EOS can be represented by the effective adiabatic index 
$\Gamma$ that is defined by $\Gamma \equiv (\rho/P) \partial P/\partial
\rho$, where $\rho$ is the rest mass density and $P$ is the pressure.
In Figure~\ref{fig:effective_gamma}, we plot $\Gamma$ against the energy 
density $\epsilon$ for the APR and FPS EOSs for comparison.  
The fact that the effective adiabatic index for the APR EOS is higher than 
that of the FPS EOS in the high-density regime above a few times the nuclear 
density $2.8\times 10^{14}\ {\rm g\ cm}^{-3}$ is the main reason why the APR 
EOS has a larger maximum-mass limit. 
To further illustrate the effect of the stiffness of the underlying EOS 
on the universal I-Love-Q relations, we repeat the analysis using a 
polytropic EOS model given by $P = K \rho^\Gamma$, where $K$ and $\Gamma$ 
are constants. We choose five different cases for the adiabatic index: 
$\Gamma = 1.6, 1.8, 2.0, 2.2$, and 2.5. 
We plot $\Delta\bar Q$ and $\Delta\bar I$ against $\ln \bar\lambda$ for these
cases in Figure~\ref{fig:i-love-q_poly}. It can clearly be seen from the 
figure that there is a monotonic trend as we vary 
the value of $\Gamma$. In particular, similar to the results for realistic 
EOSs as shown in Figure~\ref{fig:i-love-q_real}, the fractional differences
$\Delta\bar Q$ and $\Delta\bar I$ of the polytropic EOS decrease as 
$\Gamma$ is increased.

\section{Analytical study of the I-Love relation}
\label{sec:Analy}

In the previous section, we provided numerical evidence showing that  
the I-Love-Q relations of realistic EOSs and the incompressible 
model agree very well, with fractional differences of the order of 1\% 
or less for NSs spanning compactness from about 0.1 to 0.3. 
Here, we will study the I-Love relation analytically in the Newtonian limit 
and show that the incompressible limit is a crucial link to the 
observed universality of the I-Love relation.

In the Newtonian limit, the main equation for calculating the 
(quadrupolar $l=2$) tidal deformability $\lambda$ is 
(see, e.g., Equation~(60) of \citet{Yagi:2013long})
\begin{equation}
{d^2 h \over dr^2} + {2\over r}{dh \over dr}
- \left( {6\over r^2} - 4\pi \rho {d\rho\over dP} \right) h = 0 , 
\label{eq:love_h}
\end{equation}
where $\rho$ and $P$ are the mass density and pressure, respectively. 
The metric function $h$ (in the Newtonian limit) is related to the 
dimensionless tidal deformability 
$\bar \lambda=\lambda/ M^5$ by
\begin{equation}
{\bar \lambda} = { 2 - y(R) \over 3 \left[ 3+ y(R) \right] } C^{-5} , 
\label{eq:lambda_bar} 
\end{equation}
where $R$ is the radius, $C= M/R$ is the compactness, and 
$y(R) = R h^\prime(R) / h(R)$. On the other hand, the dimensionless moment 
of inertia $\bar I=I/M^3$ is 
\begin{equation}
{\bar I} = { \int_0^R \rho r^4 dr \over 24\pi^2 \left( 
\int_0^R \rho r^2 dr \right)^3 } .
\end{equation}

The starting point of our analysis is to assume that the density inside a 
compact star can be modeled well by 
\begin{equation}
\rho = \rho_0 \left( 1 - \delta x^2 \right) ,
\label{eq:rho_profile}
\end{equation}
where $\rho_0$ is the central density and $x=r/R$. The parameter $\delta$ is 
defined in the range $[0,1]$ and is used to mimic different density profiles 
of different EOS models. 
The case $\delta=0$ corresponds to the incompressible limit, 
which we will expand the I-Love relation about perturbatively.
For the case $\delta=1$, if we replace $\rho$ by the energy density 
$\epsilon$ (and similarly for $\rho_0$), then the profile is known as 
the Tolman VII model in GR and has been shown to give reasonably good 
approximations for NSs constructed with different realistic EOS models 
\citep{Lattimer:2001p426,Postnikov:2010p024016}.

\begin{figure}
  \centering
  \includegraphics*[width=7.5cm]{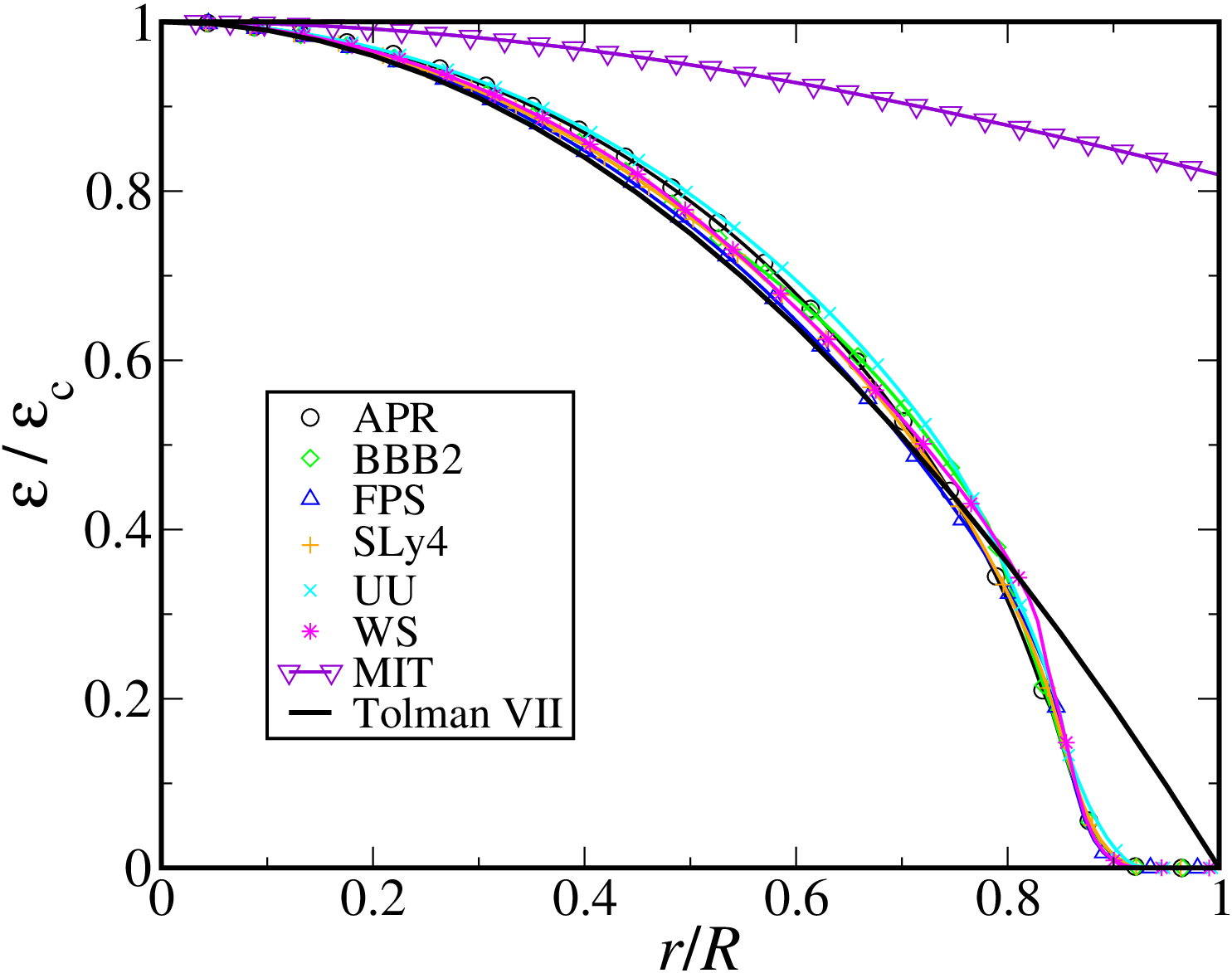}
  \caption{Energy density (normalized by its central value) is plotted
against $r/R$ for stellar models with a fixed compactness $C=0.1$.  
The solid line represents the Tolman VII model.  }
  \label{fig:dens_profile_fix_C}
\end{figure}

In Figure~\ref{fig:dens_profile_fix_C}, we plot the energy density
(normalized by the central value) against $r/R$ for NSs 
constructed from our chosen EOS models. 
All the stellar models in Figure~\ref{fig:dens_profile_fix_C} have the same 
compactness $C=0.1$. The profile given by the Tolman VII model is also plotted
in the figure for comparison. It is seen that the Tolman VII model can mimic 
the density profiles of realistic EOSs quite well. 
In particular, inside a large part of the stars, the density profiles 
of realistic EOSs are generally slightly higher than that 
of the Tolman VII model. Note also that the profile for QS is well above the 
Tolman VII profile.

In our Newtonian analysis, Equation~(\ref{eq:rho_profile}) with the parameter 
$\delta \lesssim 1$ would represent approximately the density profiles of 
NS models constructed from different realistic EOSs. 
On the other hand, for $\delta \approx 0$, the resulting density profile
would be used to mimic the structure of a QS.
Using Equation~(\ref{eq:rho_profile}), we can rewrite 
Equation~(\ref{eq:love_h}) as 
\begin{equation}
x^2 \left( 1 - {3 \over 5} \delta x^2 \right) {d^2 {\tilde h}\over dx^2}
- \left( 6 - {48\over 5} \delta x^2 \right) {\tilde h} = 0 , 
\label{eq:love_tilde_h}
\end{equation} 
where ${\tilde h} = r h$. Equation~(\ref{eq:love_tilde_h}) has a formal 
solution given by the hypergeometric function $ {}_2F_1$: 
\begin{equation}
{\tilde h} = x^3 {}_2F_1 \left( { 5 - \sqrt{65} \over 4} , 
{ 5 + \sqrt{65} \over 4 } , {7 \over 2} , {3 \over 5} \delta x^2 \right) . 
\end{equation}
The dimensionless tidal deformability $\bar \lambda$ can then be calculated 
by Equation~(\ref{eq:lambda_bar}), where the value $y(R)$ is given by 
\begin{equation}
y(R) = 2 - {6 \delta\over 7}  { {}_2F_1 \left( {9-\sqrt{65}\over 4}, 
{9+\sqrt{65}\over 4} , {9\over 2}, {3\delta\over 5} \right) \over 
{}_2F_1 \left( {5-\sqrt{65}\over 5}, {5+\sqrt{65}\over 4},{7\over 2}, 
{3\delta\over 5} \right) } - { 15 \left( 1 - \delta \right)\over 5 - 3 \delta}
 .
\label{eq:y_R} 
\end{equation}
On the other hand, the dimensionless moment of inertia $\bar I$ can be 
calculated easily and is given by 
\begin{equation} 
{\bar I} = { 2 ( 7 - 5 \delta )\over 7 (5-3\delta) } C^{-2} . 
\label{eq:I_bar} 
\end{equation}

\begin{figure}
  \centering
  \includegraphics*[width=7.5cm]{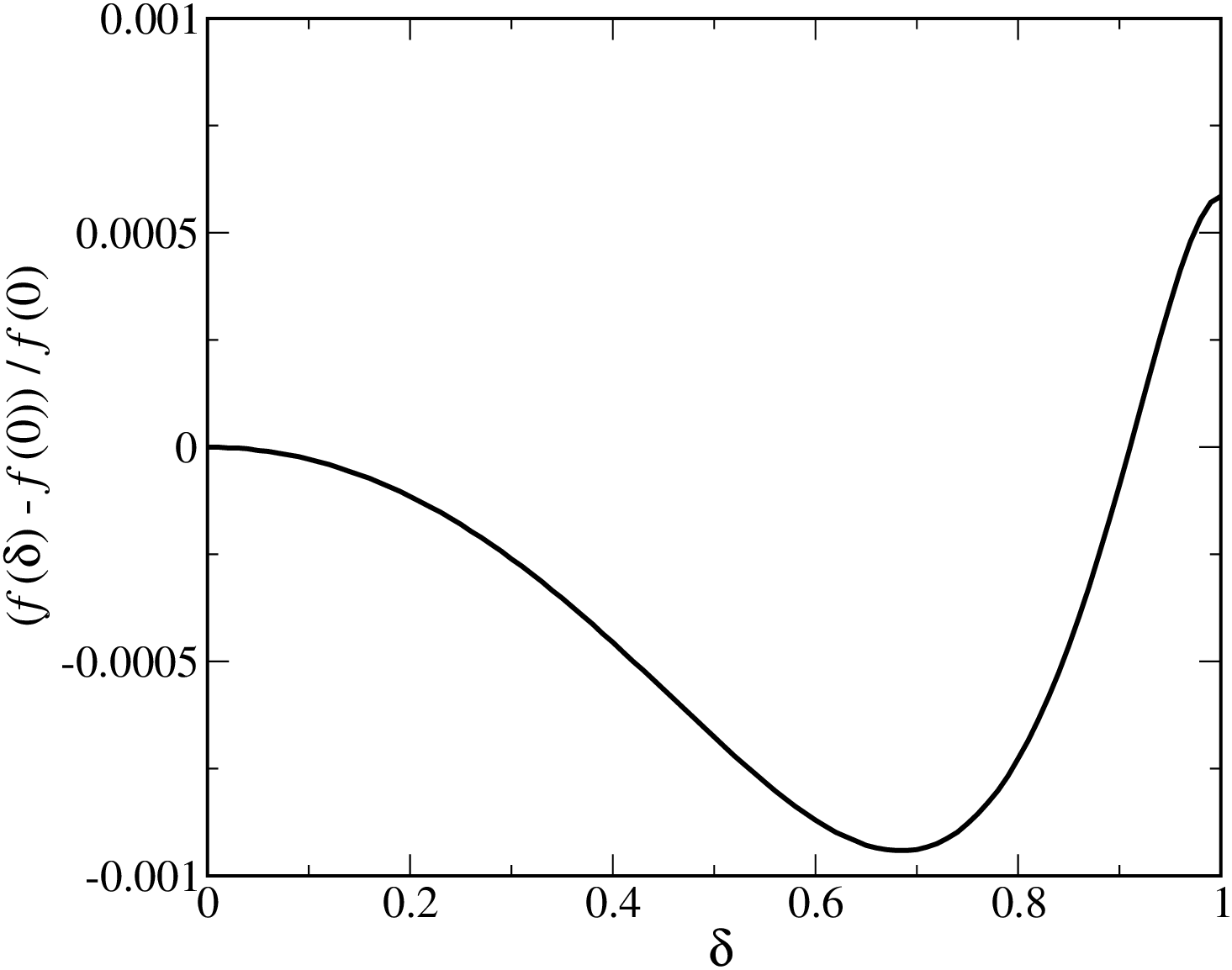}
  \caption{Variation of the I-Love relation plotted against the 
expansion parameter $\delta$.  }
  \label{fig:delta_expansion}
\end{figure}

By eliminating the compactness $C$ in Equations~(\ref{eq:lambda_bar}) and 
(\ref{eq:I_bar}), we obtain the I-Love relation connecting 
${\bar I}$ and ${\bar\lambda}$: 
\begin{equation}
f(\delta) \equiv 
{\bar \lambda} {\bar I}^{-5/2} = { 2 - y(R) \over 3( 3+y(R) )} 
\left[ { 2(7-5\delta) \over 7(5 - 3\delta) } \right]^{-5/2} ,
\label{eq:I-Love-delta}
\end{equation}
where $y(R)$ is given by Equation~(\ref{eq:y_R}). The dependence of the I-Love
relation on the parameter $\delta$, and hence implicitly the underlying EOS
model, can be seen in Figure~\ref{fig:delta_expansion}. The figure shows that 
the function $f(\delta)$ changes at most by only 0.1\% as $\delta$ 
varies from 0 to 1. This weak dependence on $\delta$ can also be shown 
analytically by expanding Equation~(\ref{eq:I-Love-delta}) perturbatively 
about the incompressible limit $\delta=0$ which gives (up to the 
second order of $\delta$) 
\begin{equation}
{\bar \lambda} {\bar I}^{-5/2} = 
5 \sqrt{5\over 2} \left( {5\over 8} - {1\over 588} \delta^2 + ... \right) ,
\label{eq:i-love-expand}
\end{equation}
where the dots inside the parentheses stand for higher-order terms of 
$\delta$. 
We note that our result reduces correctly to the I-Love relation for 
Newtonian incompressible stars when $\delta =0$ \citep{Yagi:2013long}. 
More interestingly,
it is seen that the incompressible limit is a stationary point in the sense 
that the linear-order term in $\delta$ vanishes in the expansion. 
The dependence of the I-Love relation on the expansion parameter is only 
through the second-order term in $\delta$ and hence its dependence 
on the underlying EOS is weak as long as the stellar density profile can be 
represented reasonably well by Equation~(\ref{eq:rho_profile}). 
It should be pointed out that the vanishing of the linear-order term 
in $\delta$ in Equation~(\ref{eq:i-love-expand}) is due to the fact that we 
consider the quadrupolar $l=2$ tidal deformability. For higher multipole order 
(i.e., $l>2$), the leading dependence of the I-Love relation on $\delta$ 
would generally be linear in order.

Recently, \citet{Yagi:2014p124013} pointed out that $\bar I$ and $\bar Q$ are 
dominated by a region that is 50\%-95\% of the width of the stellar radius. 
They suggested that the variation of the adiabatic index for different 
EOSs in this region is so high (of the order of 10\%) that the approximate 
similarity of EOSs in this region cannot explain the universality of the 
I-Love-Q relations. 
However, it should be emphasized that although $\bar I$ and $\bar Q$ depend 
sensitively on the underlying EOS, it is their specific combination that 
leads to the cancelation of the strong dependence on EOS, and hence the 
existence of the universality. 
This point is reflected in our analysis by noting that although $\bar I$ 
and $\bar \lambda$ depend non-trivially on the parameter $\delta$, their 
combination $\bar \lambda {\bar I}^{-5/2}$ turns out to depend weakly on 
$\delta$, hence leading to the universal I-Love relation.

\section{Breakdown of the universality}
\label{sec:breakdown}

\begin{figure}
  \centering
  \includegraphics*[width=7.5cm]{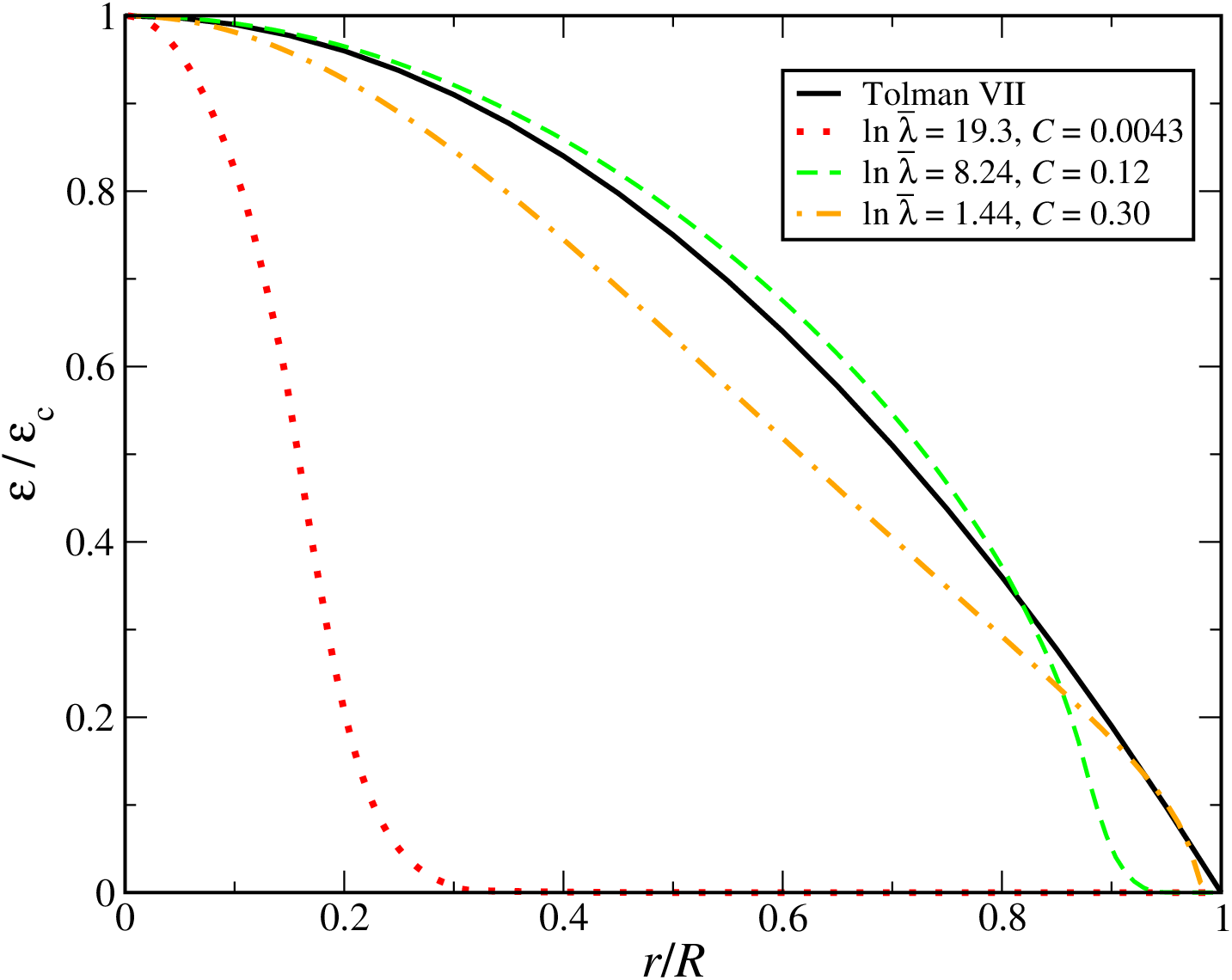}
  \caption{Energy density (normalized by its central value) is plotted
against $r/R$ for stellar models constructed by the SLy4 EOS with different 
compactness. The solid line represents the Tolman VII model.}
  \label{fig:dens_sly4_lambda}
\end{figure}

In \citet{Yagi:2013long,Yagi:2013}, it was shown analytically in the Newtonian limit that 
the I-Love-Q relations are very similar for the incompressible 
($\Gamma=\infty$) and $\Gamma=2$ polytropic stellar models, although the 
reason that this is the case was not understood. 
In the above, we have provided support for the analysis carried out 
in \citet{Yagi:2013long,Yagi:2013}. Furthermore, we have shown 
that the incompressible limit is a stationary point and the I-Love relation 
depends weakly on the parameter $\delta$, and hence on the underlying stellar 
structure and EOS. Our conclusion is valid for stellar density profiles that 
can be modeled well by Equation~(\ref{eq:rho_profile}), which essentially 
covers the region above the Tolman VII profile in 
Figure~\ref{fig:dens_profile_fix_C}. 
As we have seen, typical NSs with compactness at about 0.1 constructed from
our chosen EOS models reside in this domain of validity and the universality 
of the I-Love-Q relations in this regime is clearly demonstrated in 
Figure~\ref{fig:i-love-q_real}. In the region between $\ln \bar \lambda = 5$ 
and 10, corresponding to $C$ in the range from about 0.2 to 0.07, the 
accuracy of the universality of the I-Love-Q relations is better than 1\%. 
On the other hand, we expect (at least qualitatively) that stellar models
with density profiles that are far from and below the Tolman VII 
profile would show a much stronger dependence of the I-Love-Q relations on 
the underlying EOS models. 

\begin{figure}
  \centering
  \includegraphics*[width=7.5cm]{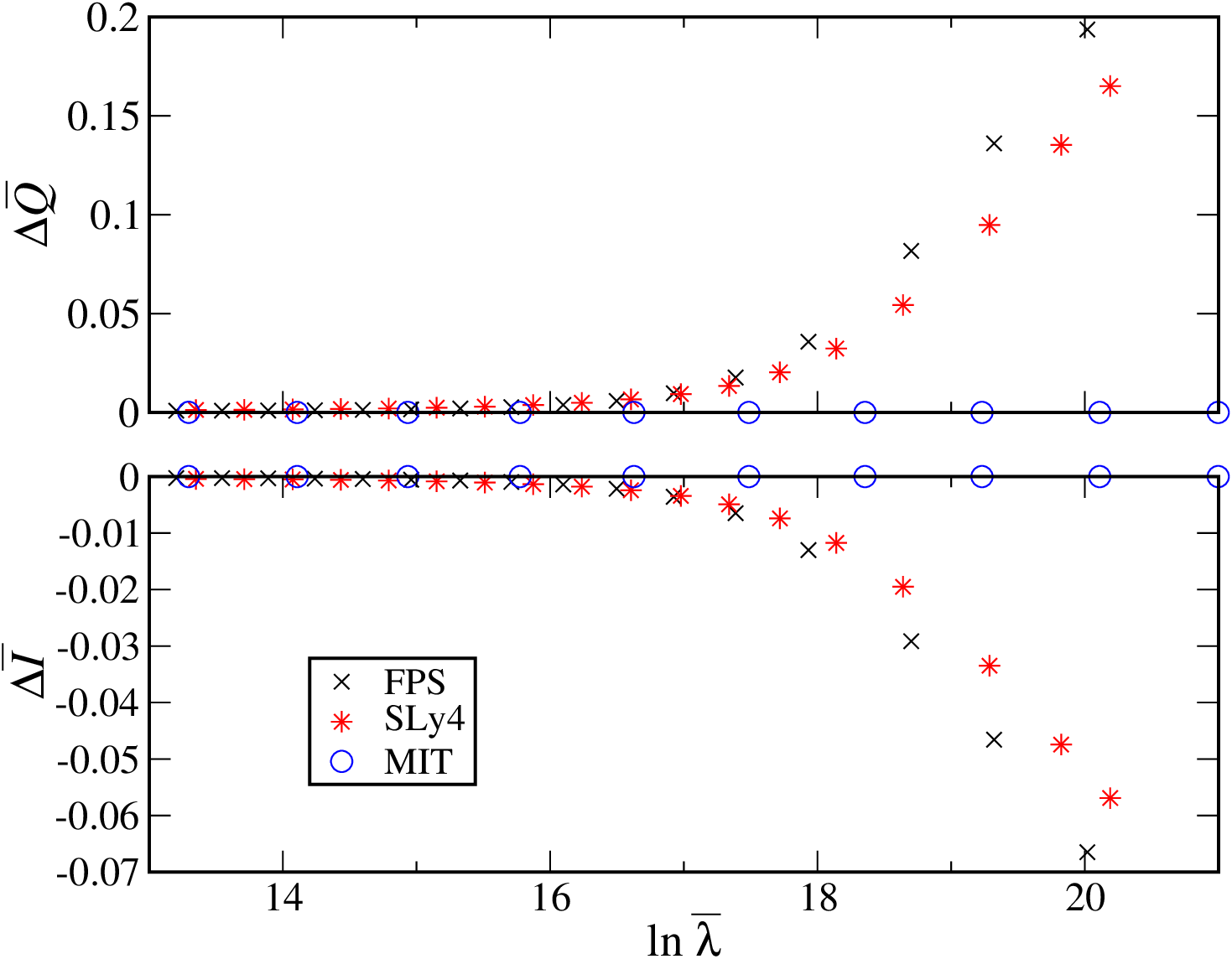}
  \caption{$\Delta \bar Q$ (upper panel) and $\Delta \bar I$ (lower panel) 
are plotted against $\ln \bar \lambda$ for low-mass NSs 
(FPS and SLy4 EOSs) and quark stars (MIT). }
  \label{fig:i-love-q_low_den}
\end{figure}

When would our analysis and the universality break down? 
Figure~\ref{fig:i-love-q_real} already shows that the dependence of the 
I-Love-Q relations on the underlying EOS model becomes more apparent in 
the ultra-relativistic regime near the maximum mass limits 
(i.e., $\ln \bar \lambda \approx 0.2$). This may not be that surprising 
because it is well known that gravity in GR is stronger than that 
in Newtonian theory, and hence GR tends to destabilize a star. 
This destabilizing effect can be interpreted as an effective softening of 
the underlying EOS in such a way that the resulting density profile 
cannot be modeled well by Equation~(\ref{eq:rho_profile}). 
This can be seen from Figure~\ref{fig:dens_sly4_lambda} where we plot the 
energy density profiles of NSs with different compactness using the SLy4 EOS. 
The Tolman VII model is also plotted for comparison (solid line). 
The profile corresponding to $\ln\bar\lambda = 8.24$ and $C=0.12$ is a 
typical NS model (dashed line) that can be modeled well by the Tolman VII 
model. This star lies in the domain where the universality of the I-Love-Q 
relations has the highest accuracy ($< 1\%$ level). On the other hand, 
the maximum-mass NS model (dashed-dotted line) has the values 
$\ln\bar\lambda=1.44$ and $C=0.30$. It can be seen that while the density 
profile of this configuration is still reasonably close to the Tolman VII 
model, it is nevertheless located below the Tolman VII profile. 
This model shows a larger deviation (e.g., a few percent for $\Delta\bar Q$) 
from the incompressible limit as shown in Figure~\ref{fig:i-love-q_real}. 

In Figure~\ref{fig:dens_sly4_lambda}, we also plot the profile for a 
$0.1M_\odot$ low-mass NS (dotted line) with $\ln\bar \lambda=19.3$ and 
$C=4.3\times 10^{-3}$. 
This model is near the minimum mass of stable NSs that can be 
supported by the SLy4 EOS. We see that the density profile
of this star cannot be modeled well by Equation~(\ref{eq:rho_profile}). 
The density profile of this model is also far away from the region of 
validity of our analysis, namely the region above the Tolman VII profile 
in the figure. 
We thus expect that the universality of the I-Love-Q relations would break 
down.
In order to check our expectation, we extend 
Figure~\ref{fig:i-love-q_real} to the low-mass region using the FPS, SLy4, 
and MIT bag models and plot the results in Figure~\ref{fig:i-love-q_low_den}. 
It can be seen that the I-Love-Q relations for low-mass NSs become 
more sensitive to the EOS as $\bar\lambda$ increases, and they also
deviate significantly from the incompressible limit.
Note, however, that the sensitivity on the EOS can only be seen 
indirectly by subtracting the I-Love-Q relations by their incompressible 
limits.
Near the minimum-mass limit at about $\ln\bar\lambda=20$, 
the fractional differences $\Delta \bar Q$ and $\Delta \bar I$ increase 
to about 20\% and 7\%, respectively. However, contrary to the case for NSs, 
the I-Love-Q relations for low-mass QSs can still be accurately
represented by the incompressible limit. 
On physical grounds, the large deviation of the I-Love-Q relations from 
the incompressible limit for low-mass NSs is due to the fact that these 
stars have low densities and their interiors are on average dominated by 
relatively softer matter with a smaller $\Gamma$. 
For instance, a low-mass NS with $\ln \bar\lambda\approx 20$ and 
$C\approx 3\times 10^{-3}$ constructed from the SLy4 EOS would have 
about 70\% of its total mass attributed to matter with $\Gamma < 2$. 
On the other hand, QSs can be modeled by the incompressible 
model very well regardless of their masses.

The different behaviors of the I-Love-Q relations for low-mass NSs and QSs 
might point to a method that allows us, at least in principle, to distinguish 
between these two types of objects observationally. 
However, can low-mass compact stars be formed in the first place? 
It is known that hot proto-NSs formed in supernova explosions generally 
have a large minimum mass, hence rendering the existence of low-mass NSs 
unlikely. Nevertheless, it is still not ruled out that low-mass NSs can be 
formed via fragmentation during the formation of proto-NSs 
\citep{Popov:2007p381, Horowitz:2010p103001}. This is still an open 
question for further investigation.

\section{Comparison with Elliptical isodensity approximation}
\label{sec:isodensity}

Approximate universal relations among multipole moments have also
been found recently
\citep{Stein:2014p15,Yagi:2014p124013,Chatziioannou:1406.7135,Yagi:1406.7587}.
In particular, \citet{Yagi:1406.7587} found that the isodensity
contours of realistic NSs can be approximated by elliptical
isodensity contours and that the relaxation of the self-similarity
assumption of these surfaces can destroy the universal relations
among multipole moments. They further suggested that the universal
I-Q relation between $\bar I$ and $\bar Q$ would be affected by
the elliptical isodensity approximation in roughly the same way.
In this paper, based on our numerical evidence and Newtonian
perturbative analysis, we propose that the universality of
I-Love-Q relations can be attributed to the incompressible limit
of these relations. The physical reason is due to the fact that
modern realistic EOSs are stiff in the nuclear density range.

Is there any connection between our study and the elliptical
isodensity approximation as discussed by \citet{Yagi:1406.7587}? A
first hint can be seen in Figure 17 of \citep{Yagi:1406.7587} in
which the variation of the eccentricity inside a slowly rotating
Newtonian star (modeled by the polytropic EOS) is shown to
decrease as the stiffness of the EOS increases. In particular, the
elliptical isodensity approximation becomes exact in the
incompressible limit. \citet{Yagi:1406.7587} also found that the
eccentricity profile is almost constant for QSs with large $\bar
I$ (corresponding to small compactness), and hence the elliptical
isodensity approximation becomes highly accurate (see Figure 15 of
\citet{Yagi:1406.7587}). We can understand this result by noting
that the average effective adiabatic index of quark matter inside
QSs increases with decreasing compactness. On the other hand,
\citet{Yagi:1406.7587} found that there are no universal relations
among multipole moments and between ${\bar I}$ and ${\bar Q}$ for
noncompact stars.\footnote{Since the EOS for noncompact stars such
as the sun is well understood, \citet{Yagi:1406.7587} used
different opacity laws to mimic the effects of different EOS
models in their work.} They also found that the variation of
eccentricity inside noncompact stars is large and suggested that
the loss of universality for these stars could be a consequence of
the breakdown of the elliptical isodensity approximation. From our
point of view, however, the loss of universality for noncompact
stars is due to the fact that the underlying EOS is much softer
than the EOS of nuclear matter.

Generally speaking, for a given  star, the validity of the
elliptical isodensity approximation, which is a criterion for the
I-Love-Q universality as suggested in
\citet{Stein:2014p15}, \citet{Chatziioannou:1406.7135}, 
\citet{Yagi:1406.7587}, needs to be verified by numerical methods on a case 
by case basis. However,
from the above discussion, it is evident that the high stiffness
of the EOS can directly lead to the validity of the elliptical
isodensity approximation.  Therefore, we believe that the
stiffness of the EOS plays a more fundamental role in the observed
universality. The requirement that nuclear-matter EOSs only need to 
be stiff is a physically appealing and simple explanation for the 
universality.

On the other hand,  \citet{Yagi:1406.7587} further suggested that
the validity of the elliptical isodensity approximation and hence
the observed universality, are due to an  approximate symmetry
emerging in relativistic stars, which are composed of nuclear
matter with high stiffness and which have large compactnesses. It is
crucial to pinpoint the exact cause of the universality. Does
either high stiffness or large compactness in itself lead to the
universality? Does the universality arise from the concerted
effects of both high stiffness and large compactness? 
As we have already seen in Figures~\ref{fig:i-love-q_real} and 
\ref{fig:i-love-q_poly} of the present paper, the accuracy of the 
I-Love-Q relations slightly deteriorates towards the large compactness 
end and, as mentioned above, is better for high-stiffness EOSs. 
Similarly, as shown in Figures 2, 15, and 16 of \citet{Yagi:1406.7587}, 
the accuracy of the elliptical isodensity approximation also worsens 
with increasing stellar compactness.
Therefore, we conclude that the high stiffness of nuclear matter
is the cause of the universality. The effect of GR in fact softens
the stiffness of matter and adversely affects the universal
relations.

\section{Conclusion}
\label{sec:conclude}

\begin{figure}
  \centering
  \includegraphics*[width=7.5cm]{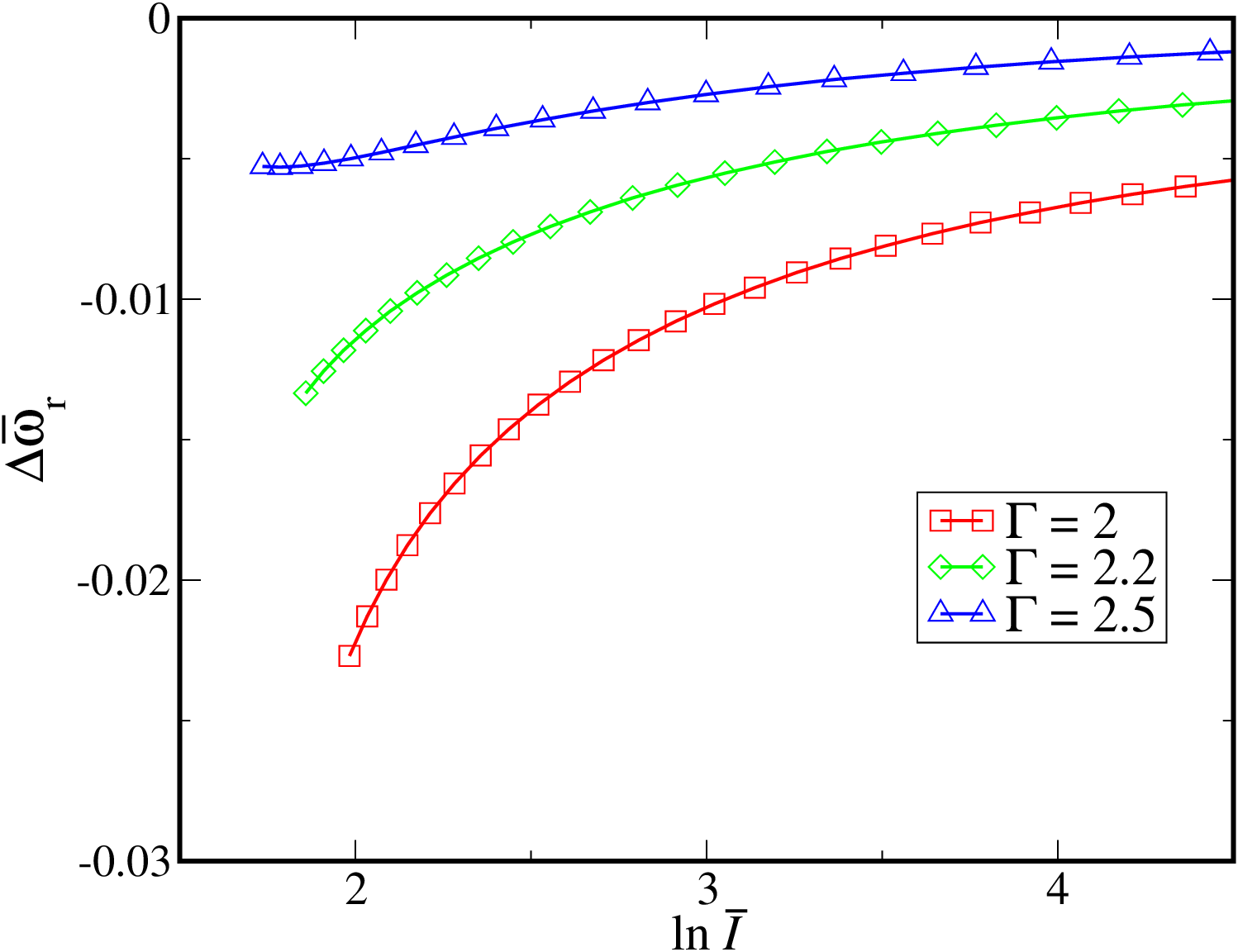}
  \caption{$\Delta \bar \omega_{\rm r}$ is plotted against $\ln \bar I$ for 
polytropic models.}
  \label{fig:fmode_i_poly}
\end{figure}

In this paper, we have provided numerical evidence showing that the 
universality of I-Love-Q relations can be attributed to the incompressible 
limits of these relations.  
For typical NSs with compactness in the range from about 0.05 to 0.2, 
the incompressible limit can model the I-Love-Q relations of
realistic EOSs to much better than 1\% level. 
Furthermore, using a generalized Tolman VII model density profile,
we have carried out a perturbative analysis of the I-Love 
relation in Newtonian gravity and have shown that the relation can be 
represented by the incompressible limit accurately and is weakly dependent 
on the EOS because the incompressible limit is a stationary point. 
The leading dependence of the I-Love relation on the density profile, 
and hence the underlying EOS model, turns out to be second order in the 
expansion parameter $\delta$ that is used to mimic realistic NSs 
(when $\delta \approx 1$) and QSs (when $\delta \approx 0$).
We also demonstrate numerically that the I-Love-Q relations for low-mass
NSs, which are composed mainly of softer nuclear matter,  
deviate significantly from the incompressible limit and become more 
sensitive on the underlying EOS. 
On the other hand, the I-Love-Q relations for low-mass QSs can still be 
represented accurately by the incompressible limit because QSs can be modeled 
well by the incompressible stellar model regardless of their masses.

One of the possible reasons for the universality of I-Love-Q relations,
as suggested in \citet{Yagi:2013long,Yagi:2013}, is that these relations 
approach the black hole limit as the NS compactness increases. 
In this work, we show that there exists a ``softer'' limit, the 
incompressible limit, to which the I-Love-Q relations for different 
realistic EOSs converge, and hence a universality is established. 
Besides the I-Love-Q relations, it is also known that the $f$-mode 
oscillation frequency and moment of inertia of compact stars also display 
a universality \citep{Lau:2010p1234}. 
It would be interesting to study whether the two 
different sets of $f$-mode and I-Love-Q universal relations have a common 
origin or not. A first hint can be seen in Figure~\ref{fig:fmode_i_poly}, in 
which we plot the fractional difference for the real part of the 
$f$-mode oscillation frequency $\Delta \bar \omega_{\rm r}$ 
(defined similarly as $\Delta \bar I$ in Equation~(\ref{eq:delta_I})) against 
$\ln \bar I$ for three different polytropic EOS models, where 
$\bar \omega_{\rm r}$ is the scaled frequency $M\omega_{\rm r}$. 
We see that the trend of the results is essentially 
the same as those in Figure~\ref{fig:i-love-q_poly} for the I-Love-Q 
relations. The incompressible limit can generally model the $f$-mode 
universal relation to a percent level. This strongly suggests that the 
incompressible limit also plays an important role in the $f$-mode universal
relations. We conjecture that in view of the $f$-mode and I-Love-Q 
relations, which is supported by our numerical evidence, their universalities
originate from the fact that modern realistic EOSs turn out to 
be stiff enough to be modeled well by the incompressible limit of these 
relations.




\begin{thebibliography}{}

\bibitem[{Akmal {et~al.}(1998)Akmal, Pandharipande, \&
  Ravenhall}]{Akmal:98p1804}
Akmal, A., Pandharipande, V.~R., \& Ravenhall, D.~G. 1998, PhRvC, 58,
  1804

\bibitem[{Alcock {et~al.}(1986)Alcock, Farhi, \& Olinto}]{Alcock:1986p261}
Alcock, C., Farhi, E., \& Olinto, A. 1986, ApJ, 310, 261

\bibitem[{Andersson \& Kokkotas(1996)}]{Andersson:1996p20}
Andersson, N., \& Kokkotas, K.~D. 1996, PhRvL, 77, 20

\bibitem[{Andersson \& Kokkotas(1998)}]{Andersson:1998p1059}
---. 1998, MNRAS, 299, 1059

\bibitem[{Antoniadis {et~al.}(2013)}]{Antoniadis:2013p6131}
Antoniadis, J., Freire, P.~C.~C., Wex, N., {et~al.} 2013, Sci, 340, 6131 

\bibitem[{Baldo {et~al.}(1997)Baldo, Bombaci, \& Burgio}]{Baldo:1997p274}
Baldo, M., Bombaci, I., \& Burgio, G.~F. 1997, A \& A, 328, 274

\bibitem[{Bejger \& Haensel(2002)}]{Bejger:2002p8392}
Bejger, M., \& Haensel, P. 2002, A \& A, 396, 917

\bibitem[{Benhar {et~al.}(1999)Benhar, Berti, \& Ferrari}]{Benhar1999:p797}
Benhar, O., Berti, E., \& Ferrari, V. 1999, MNRAS, 310, 797

\bibitem[{Benhar {et~al.}(2004)Benhar, Ferrari, \&
  Gualtieri}]{Benhar:2004:p124015}
Benhar, O., Ferrari, V., \& Gualtieri, L. 2004, PhRvD, 70, 124015

\bibitem[{Binnington \& Poisson(2009)}]{Binnington:09p084018}
Binnington, T., \& Poisson, E. 2009, PhRvD, 80, 084018

\bibitem[{{Chakrabarti} {et~al.}(2014){Chakrabarti}, {Delsate},
  {G{\"u}rlebeck}, \& {Steinhoff}}]{Chakrabarti:1311.6509}
{Chakrabarti}, S., {Delsate}, T., {G{\"u}rlebeck}, N., \& {Steinhoff}, J. 2014,
  PhRvL, 112, 201102

\bibitem[{{Chatziioannou} {et~al.}(2014){Chatziioannou}, {Yagi}, \&
  {Yunes}}]{Chatziioannou:1406.7135}
{Chatziioannou}, K., {Yagi}, K., \& {Yunes}, N. 2014, PhRvD, 90, 064030

\bibitem[{Damour \& Nagar(2009)}]{Damour:09p084035}
Damour, T., \& Nagar, A. 2009, PhRvD, 80, 084035

\bibitem[{Demorest {et~al.}(2010)}]{Demorest:2010p1081}
Demorest, P.~B., Pennucci, T., Ransom, S.~M., Roberts, M.~S.~E., \&
Hessels, J.~W.~T. 2010, Natur, 467, 1081


\bibitem[{{Doneva} {et~al.}(2014a){Doneva}, {Yazadjiev}, {Staykov}, \&
  {Kokkotas}}]{Doneva:1408.1641}
{Doneva}, D.~D., {Yazadjiev}, S.~S., {Staykov}, K.~V., \& {Kokkotas}, K.~D.
  2014b, arXiv:1408.1641 [gr-qc]

\bibitem[{{Doneva} {et~al.}(2014b){Doneva}, {Yazadjiev}, {Stergioulas}, \&
  {Kokkotas}}]{Doneva:1310.7436}
{Doneva}, D.~D., {Yazadjiev}, S.~S., {Stergioulas}, N., \& {Kokkotas}, K.~D.
  2014a, \apjl, 781, L6


\bibitem[{Douchin \& Haensel(2000)}]{Douchin:2000p107}
Douchin, F., \& Haensel, P. 2000, PhLB, 485, 107

\bibitem[{Flanagan \& Hinderer(2008)}]{Flanagan:08p021502}
Flanagan, E.~E., \& Hinderer, T. 2008, PhRvD, 77, 021502

\bibitem[{Hartle(1967)}]{Hartle:67p1005}
Hartle, J.~B. 1967, \apj, 150, 1005

\bibitem[{Hartle \& Thorne(1968)}]{Hartle:68:p907}
Hartle, J.~B., \& Thorne, K.~S. 1968, \apj, 153, 807

\bibitem[{{Haskell} {et~al.}(2014){Haskell}, {Ciolfi}, {Pannarale}, \&
  {Rezzolla}}]{Haskell:1309.3885}
{Haskell}, B., {Ciolfi}, R., {Pannarale}, F., \& {Rezzolla}, L. 2014, \mnras,
  438, L71

\bibitem[{Hinderer(2008)}]{Hinderer:08p1216}
Hinderer, T. 2008, \apj, 677, 1216

\bibitem[{{Horowitz}(2010)}]{Horowitz:2010p103001}
{Horowitz}, C.~J. 2010, PhRvD, 81, 103001

\bibitem[{{Lattimer}(2012)}]{Lattimer:2012p485}
{Lattimer}, J.~M. 2012, ARNPS, 62, 485

\bibitem[{Lattimer \& Prakash(2001)}]{Lattimer:2001p426}
Lattimer, J.~M., \& Prakash, M. 2001, \apj, 550, 426

\bibitem[{Lattimer \& Schutz(2005)}]{Lattimer:2005p7082}
Lattimer, J.~M., \& Schutz, B.~F. 2005, \apj, 629, 979

\bibitem[{Lau {et~al.}(2010)Lau, Leung, \& Lin}]{Lau:2010p1234}
Lau, H.~K., Leung, P.~T., \& Lin, L.~M. 2010, \apj, 714, 1234

\bibitem[{Lorenz {et~al.}(1993)Lorenz, Ravenhall, \& Pethick}]{Lorenz:93p379}
Lorenz, C.~P., Ravenhall, D.~G., \& Pethick, C.~J. 1993, PhRvL, 70, 379

\bibitem[{{Maselli} {et~al.}(2013){Maselli}, {Cardoso}, {Ferrari}, {Gualtieri},
  \& {Pani}}]{Maselli:2013}
{Maselli}, A., {Cardoso}, V., {Ferrari}, V., {Gualtieri}, L., \& {Pani}, P.
  2013, PhRvD, 88, 023007

\bibitem[{{Pani} \& {Berti}(2014)}]{Pani:1405.4547}
{Pani}, P., \& {Berti}, E. 2014, PhRvD, 90, 024025 

\bibitem[{{Pappas} \& {Apostolatos}(2014)}]{Pappas:1311.5508}
{Pappas}, G., \& {Apostolatos}, T.~A. 2014, PhRvL, 112, 121101

\bibitem[{{Popov} {et~al.}(2007){Popov}, {Blaschke}, {Grigorian}, \&
  {Prokhorov}}]{Popov:2007p381}
{Popov}, S., {Blaschke}, D., {Grigorian}, H., \& {Prokhorov}, M. 2007, \apss,
  308, 381

\bibitem[{{Postnikov} {et~al.}(2010){Postnikov}, {Prakash}, \&
  {Lattimer}}]{Postnikov:2010p024016}
{Postnikov}, S., {Prakash}, M., \& {Lattimer}, J.~M. 2010, PhRvD, 82, 024016

\bibitem[{{Sham} {et~al.}(2014){Sham}, {Lin}, \& {Leung}}]{Sham:2014p66}
{Sham}, Y.-H., {Lin}, L.-M., \& {Leung}, P.~T. 2014, \apj, 781, 66


\bibitem[{{Stein} {et~al.}(2014){Stein}, {Yagi}, \& {Yunes}}]{Stein:2014p15}
{Stein}, L.~C., {Yagi}, K., \& {Yunes}, N. 2014, \apj, 788, 15

\bibitem[{Tsui \& Leung(2005{\natexlab{a}})}]{Tsui:2005p151101}
Tsui, L.~K., \& Leung, P.~T. 2005{\natexlab{a}}, PhRvL, 95, 151101

\bibitem[{Tsui \& Leung(2005{\natexlab{b}})}]{Tsui:2005p1029}
---. 2005{\natexlab{b}}, MNRAS, 357, 1029

\bibitem[{Urbanec {et~al.}(2013)Urbanec, Miller, \&
  Stuchl\'{i}k}]{Urbanec:13p1903}
Urbanec, M., Miller, J.~C., \& Stuchl\'{i}k, Z. 2013, MNRAS, 433, 1903

\bibitem[{Wiringa {et~al.}(1988)Wiringa, Fiks, \& Fabrocini}]{Wiringa:98p1010}
Wiringa, R.~B., Fiks, V., \& Fabrocini, A. 1988, PhRvC, 38, 1010

\bibitem[{Witten(1984)}]{Witten:1984p272}
Witten, E. 1984, PhRvD, 30, 272

\bibitem[{{Yagi} {et~al.}(2014{\natexlab{a}}){Yagi}, {Kyutoku}, {Pappas},
  {Yunes}, \& {Apostolatos}}]{Yagi:2014p124013}
{Yagi}, K., {Kyutoku}, K., {Pappas}, G., {Yunes}, N., \& {Apostolatos}, T.~A.
  2014{\natexlab{a}}, PhRvD, 89, 124013

\bibitem[{{Yagi} {et~al.}(2014{\natexlab{b}}){Yagi}, {Stein}, {Pappas},
  {Yunes}, \& {Apostolatos}}]{Yagi:1406.7587}
{Yagi}, K., {Stein}, L.~C., {Pappas}, G., {Yunes}, N., \& {Apostolatos}, T.~A.
  2014{\natexlab{b}}, PhRvD, 90, 063010


\bibitem[{{Yagi} \& {Yunes}(2013{\natexlab{a}})}]{Yagi:2013long}
{Yagi}, K., \& {Yunes}, N. 2013{\natexlab{a}}, PhRvD, 88, 023009

\bibitem[{{Yagi} \& {Yunes}(2013{\natexlab{b}})}]{Yagi:2013}
---. 2013{\natexlab{b}}, Sci, 341, 365

\end{thebibliography}
\end{document}